\documentclass[times]{cpeauth}

\usepackage{moreverb}

\usepackage{color}
\usepackage{soul}
\soulregister\cite7
\soulregister\section7
\soulregister\subsection7
\soulregister\item7
\soulregister\textbf7

\usepackage[colorlinks,bookmarksopen,bookmarksnumbered,citecolor=red,urlcolor=red]{hyperref}
\usepackage{rotating}

\newcommand\BibTeX{{\rmfamily B\kern-.05em \textsc{i\kern-.025em b}\kern-.08em
T\kern-.1667em\lower.7ex\hbox{E}\kern-.125emX}}

\def\volumeyear{2010}

\begin{document}


\runningheads{M.Xu et al.}{A Survey on Load Balancing Algorithms for VM Placement in Cloud Computing}

\title{A Survey on Load Balancing Algorithms for Virtual Machines Placement in Cloud Computing}

\author{Minxian Xu\affil{1}\corrauth, Wenhong Tian \affil{2}\affil{,3}, Rajkumar Buyya\affil{1}}


\address{\affilnum{1}Cloud Computing and Distributed Systems (CLOUDS) Labratory, School of Computing and Information Systems, The University of Melbourne, Australia\break
	\affilnum{2}School of Information and Software Engineering, University of Electronic Science and Technology of China, Chengdu, China\break
	\affilnum{3}Chongqing Institute of Green and Intelligent Technology, Chinese Academy of Science, Chongqing, China}

\corraddr{School of Computing and Information Systems, Doug McDonell Building, The University of Melbourne, Parkville 3010, VIC, Australia. E-mail: xianecisp@gmail.com}

\begin{abstract}
The emergence of cloud computing based on virtualization technologies brings huge opportunities to host virtual resource at low cost without the need of owning any infrastructure.  Virtualization technologies enable users to acquire, configure and be charged on pay-per-use basis. However, Cloud data centers mostly comprise heterogeneous commodity servers hosting multiple virtual machines (VMs) with potential various specifications and fluctuating resource usages, which may cause imbalanced resource utilization within servers that may lead to performance degradation and  service level agreements (SLAs) violations. To achieve efficient scheduling, these challenges should be addressed and solved by using load balancing strategies, which have been proved to be NP-hard problem. From multiple perspectives, this work identifies the challenges and analyzes existing algorithms for allocating VMs to hosts in infrastructure Clouds, especially focuses on load balancing. A detailed classification targeting load balancing algorithms for VM placement in cloud data centers is investigated and the surveyed algorithms are classified according to the classification. The goal of this paper is to provide a comprehensive and comparative understanding of existing literature and aid researchers by providing an insight for potential future enhancements.   

\end{abstract}


\maketitle


\section{Introduction}

In traditional data centers, applications are tied to specific physical servers that are often over-provisioned to deal with the upper-bound workload. Such configuration makes data centers expensive to maintain with wasted energy and floor space, low resource utilization and significant management overhead. With virtualization technology, Cloud data centers become more flexible, secure and provide better support for on-demand allocation. It hides server heterogeneity, enables server consolidation, and improves server utilization \cite{Daniels}\cite{Speitkamp}. A host is capable of hosting multiple VMs with potential different resource specifications and variable workload types. Servers hosting heterogeneous VMs with variable and unpredictable workloads may cause a resource usage imbalance, which results in performance deterioration and violation of service level agreements (SLAs) \cite{Gutierrez}. Imbalance resource usage \cite{Kerr} can be observed in cases, such as a VM is running a computation-intensive application while with low memory requirement. 

Cloud data centers are highly dynamic and unpredictable due to 1) irregular resource usage patterns of consumers constantly requesting VMs, 2) fluctuating resource usages of VMs, 3) unstable rates of arrivals and departure of data center consumers, and 4) the performance of hosts when handling different load levels may vary greatly. These situations are easy to trigger unbalanced loads in cloud data center, and they may also lead to performance degradation and service level agreement violations, which requires a load balancing mechanism to mitigate this problem. 


Load balancing in clouds is a mechanism that distributes the excess dynamic local workload ideally-balanced across all the nodes \cite{Randles}. It is applied to achieve both better user satisfaction and higher resource utilization, ensuring that no single node is overwhelmed, thus improving the system overall performance. For VM scheduling with load balancing objective in cloud computing, it aims to assign VMs to suitable hosts and balance the resource utilization within all of the hosts. Proper load balancing algorithms can help in utilizing the available resources optimally, thereby minimizing the resource consumption. It also helps in implementing fail-over, enabling scalability, avoiding bottlenecks and over-provisioning and reducing response time \cite{Kansal}.   Fig. 1 shows the application, VM and host relationship in cloud data centers. The hosts at the bottom represent the real resource for provisions, like CPU, memory and storage resource. Upper the hosts, the server virtualization platform like XEN, makes the physical resource be virtualized and manages the VMs hosted by hosts. The applications are executed on VMs and may have predefined dependencies between them. Each host could be allocated with multiple VMs, and VMs are installed with multiple applications. Load balancing algorithms are applied both at the application level and the VM level. At the application level, the load balancing algorithm is integrated into Application Scheduler, and at the VM level, the load balancing algorithm can be integrated into VM Manager. This survey paper mainly focuses on the load balancing algorithms at VM level to improve hosts performance, which is often modeled as bin packing problem and has been proved as NP-hard problem \cite{Coffman}. \\

\begin{figure}
	\centering
	\includegraphics[width=0.7\textwidth, angle=-0] {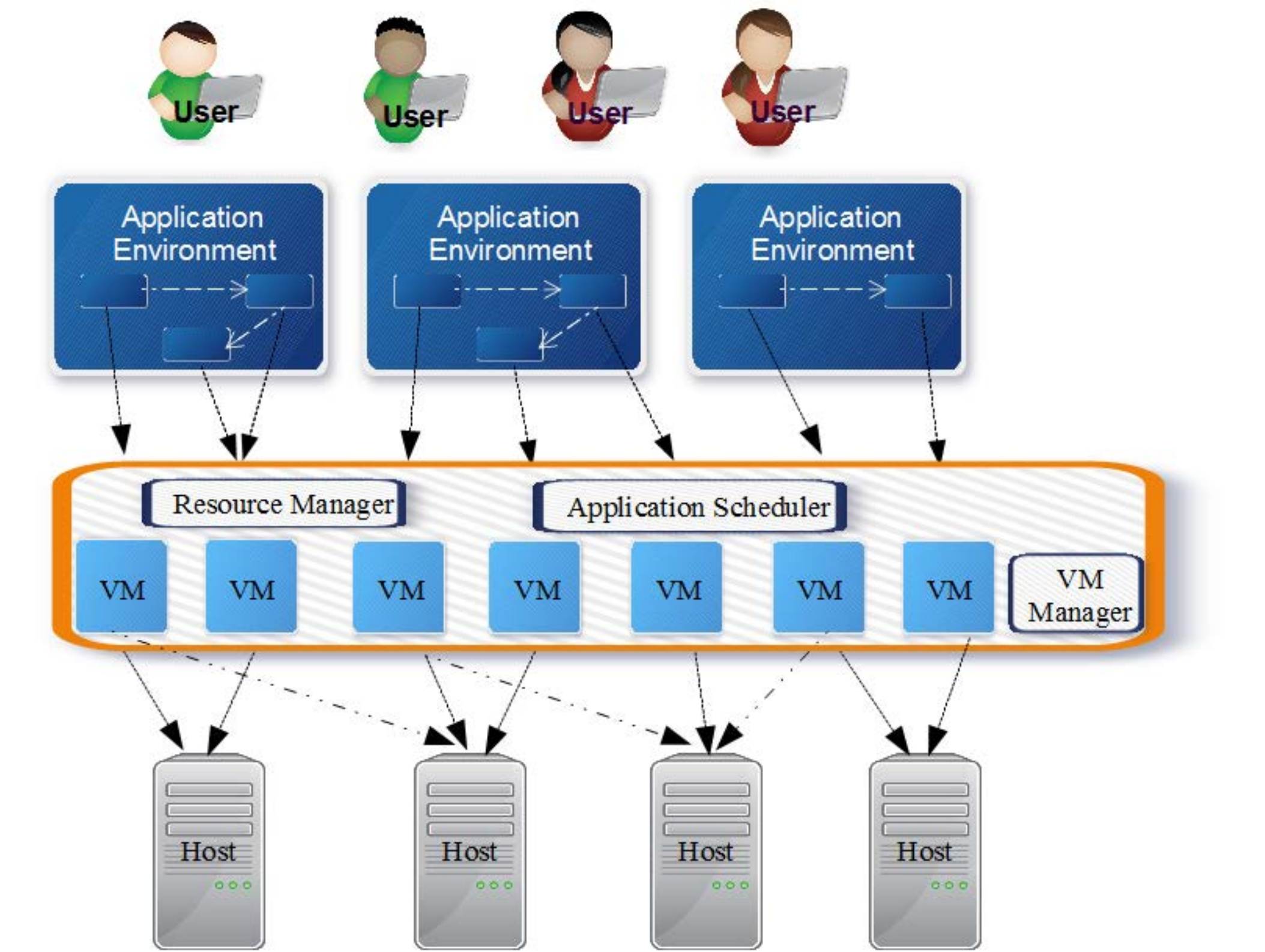}
	\caption{Application, VM and Host relationship in Cloud Data Center}
\end{figure}

The challenges of load balancing algorithms for VM placement \footnote{we note \textbf{load balancing algorithms for VM placement} as \textbf{VM load balancing algorithms} in the following sections} on host lies in follows:

\textbf{Overhead:} It determines the amount of overhead involved while implementing a load balancing system. It is composed of overhead due to VM migration cost or communication cost. A well-designed load balancing algorithm should reduce overhead.

\textbf{Performance:}  It is defined as the efficiency of the system. Performance can be indicated from users’ experience and satisfaction. How to ensure performance is a considerate challenge for VM load balancing algorithms. The performance includes following perspectives:

1) Resource Utilization: It is used to measure whether a host is overloaded or underutilized. According to different VM load balancing algorithms, overloaded hosts with higher resource utilization should be offloaded. 

2) Scalability: It represents that the quality of service keeps smooth, even if the number of users increases, which is associated with algorithm management approach, like centralized or distributed.

3) Response Time: It can be defined as the amount of time taken to react by a load balancing algorithm in a cloud system. For better performance, this parameter should be reduced. 

\textbf{The point of Failure:} It is designed to improve the system in such a way that the single point failure does not affect the provisioning of services. Like in centralized system, if one central node fails, then the whole system would fail, so load balancing algorithms should be designed to overcome this problem. \\

In this survey, we extend and complement the classifications from existing survey works through analyzing the different characteristics for VM load balancing comprehensively, like the scheduling scenario, management approaches, resource type, VM type uniformity and allocation dynamicity. We also summarize the scheduling metrics for VM load balancing algorithms, and these metrics could be used to evaluate the load balancing effects as well as other additional scheduling objectives. We then discuss performance evaluation approaches followed by existing work, which show the popular realistic platforms and simulation toolkits for researching VM load balancing algorithms in Clouds. Through a detailed discussion of existing VM load balancing algorithms, the strength and weakness of different algorithms are also presented in this survey. 

The rest of the paper is organized as follows: Section 2 introduces the related technology for VM load balancing and the general VM load balancing scenarios as well as management approaches. Section 3 discusses models for VM load balancing, including VM resource type, VM type uniformity, VM dynamicity, and scheduling process while Section 4 presents different scheduling metrics of load balancing algorithms. Section 5 compares different algorithms from implementation and evaluation perspective. Detailed introductions for a set of VM load balancing algorithms are summarized in Section 6. Finally, challenges and future directions are given in Section 7.

\section{VM Load Balancing Scenario and Management}

\subsection{Related Work}
Although there are some survey papers related to this topic, they are partially focused on VM load balancing.  
Jiang \cite{Jiang} summarized the general characteristics of distributed systems and studied task allocation and load balancing in these systems. However, this paper has not focused on cloud environment and not relevant to VM scheduling. 
Mann et al. \cite{Mann} proposed a comprehensive survey of the state-of-the-art on VM allocation in cloud data centers with a more general view. They discussed the VM allocation problem based on models and algorithmic approaches and gave algorithm suggestions for different scenarios.
However, this survey is also not concentrating on VM load balancing perspective.  
In \cite{Milani}, load balancing algorithms in Clouds were detailed classified and several algorithms were discussed with both advantages and disadvantages. This paper also addressed the challenges of these discussed algorithms. However, the discussed algorithms are not applied to VMs.  
Tiwan et al. \cite{Tiwari} gave a brief introduction for several load balancing algorithms while their limitations are not discussed, and these algorithms are also simply classified as dynamic and static ones. 
Khiyaita et al. \cite{Khiyaita} provided an overview of load balancing in Clouds and outlined the main challenges, while only limited comparisons of four load balancing algorithms were analyzed. 
Mesbahi et al. \cite{Mesbahi} evaluated three load balancing algorithms for Clouds under simulated environment and gave recommendations for different combinations.
In our survey, we concentrated on VM load balancing algorithms and complemented the classifications from existing surveys through comprehensive analysis of VM load balancing algorithms from multiple aspects, including platform type, QoS constraints, migration approach and cost, scheduling scalability and objective.

\subsection{Related Technology}
Before we discuss the VM load balancing algorithms, we firstly introduce some related technologies for load balancing. 

\textbf{Virtualization technology:} Virtualization reinforces the ability and capacity of existing infrastructure and resource, and opens opportunities for cloud data centers to host applications on shared infrastructure. VM technology was firstly introduced in the 1960s and has been widely exploited in recent years for consolidating hardware infrastructure in enterprise data centers with technologies like VMware \cite{VMware} and Xen \cite{Singh}.

\textbf{VM Migration: }Live migration of virtual machines \cite{Clark} means the virtual machine seems to be responsive all the time during the migration process from the user’ perspective. Compared with traditional suspend/resume migration, live migration brings many benefits such as energy saving, load balancing and online maintenance \cite{Ye}. Voorsluys et al \cite{Voorsluys} evaluate the VM live migration effects on the performance of applications running inside Xen VMs and show the results that migration overhead is acceptable, but cannot be disregarded.  Since the live migration technology is widely supported in the current cloud computing data center, live migration of multiple virtual machines becomes a common activity.

\textbf{VM Consolidation:} The VM consolidation is also implemented in cloud computing depending on the resource requirements of VMs. VM consolidation increases the number of suspended servers and performs VM live migration. This also helps in implementing fault tolerance by migrating the VMs from failure.

\subsection{Scenario}
 We outline the scenarios for VM load balancing algorithms as public cloud, private cloud and hybrid cloud. Under different scenarios, the algorithms may have different constraints. 

\textbf{Public Cloud:} The public cloud refers to when a Cloud is made available in a pay-as-you-go manner \cite{Armbrust}.  Several key benefits to service providers are offered by the public cloud, including no initial capital investment on infrastructure and shifting of risks to infrastructure providers. However, public clouds lack fine-grained control over data, network and security settings, which hampers their effectiveness in many business scenarios \cite{Zhao1}. Due to the lack of standardization, various and frequently changing APIs make it difficult to capture all the VMs and hosts information in this scenario.  Moreover, unpredictable load or periodical load is another challenge for VM load balancing algorithms. Therefore, some research has adopted historic data to predict future load to overcome this challenge \cite{Hu}\cite{Wen}.

\textbf{Private Cloud:} The Private Cloud term refers to internal datacenters of a business or other organization not made available to the general public. Although a public Cloud has the benefit of reduced capital investment and better deployment speed, private Clouds are even more popular among enterprises according to a survey by IDG in \cite{IDG}. The survey revealed that companies tend to optimize existing infrastructure with the implementation of a private Cloud which results in a lower total cost of ownership. In some academic experiments, the private clouds with mini size are implemented to evaluate VM load balancing performance. As within private cloud, more complex load balancing algorithms could be deployed and tested by defining more constraints like limiting the number of migrations. Compared to the public cloud, the loads are comparatively predicted and controlled, so heuristic algorithms like Ant Colony Optimization and Particle Swarm Optimization could be applied. An example of the private cloud is the intra-cloud network that connects a customer’s instances among themselves and with the shared services offered by a cloud. Within a cloud, the intra-datacenter network often has quite different properties compared to the inter-datacenter network \cite{LiAng}. Therefore, dealing with the VM load balancing problem in a private cloud, the performance like throughput would be considered as a constraint.

\textbf{Hybrid clouds:} A hybrid cloud is a combination of public and private cloud models that tries to address the limitations of each approach. In a hybrid cloud, part of the service infrastructure runs in private clouds while the remaining part runs in public clouds. Hybrid clouds offer more flexibility than both public and private clouds. Specifically, they provide tighter control and security over application data compared to public clouds, while still facilitating on-demand service expansion and contraction. On the downside, designing a hybrid cloud requires carefully determining the best split between public and private cloud components \cite{ZhangQi}. Under this condition, the communication cost would be the main constraint for VM load balancing algorithms. For instance, in a distributed cloud, requests may have the constraint that these requests are required to be allocated to a specific data center. In addition, in a multi-cloud that involves two or more clouds (public cloud and private cloud) \cite{Dana}, the migrations operations may be related to load migration from a private cloud to a public cloud. 

\subsection{Centralized and Distributed Management}
\begin{figure}
	\centering
	\includegraphics[width=0.7\textwidth, angle=-0] {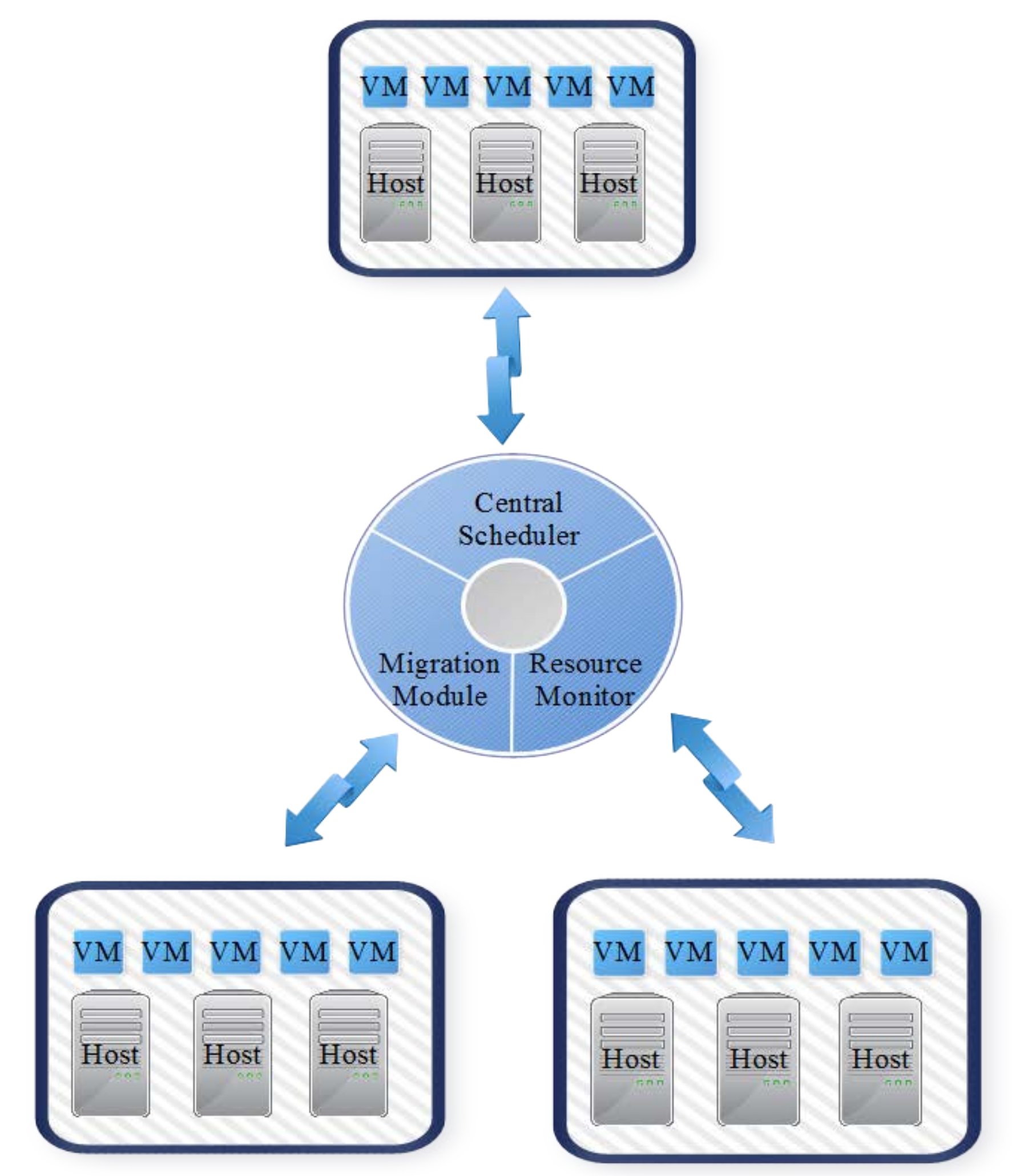}
	\caption{Centralized Scheduler}
\end{figure}

\begin{figure}
	\centering
	\includegraphics[width=0.9\textwidth, angle=-0] {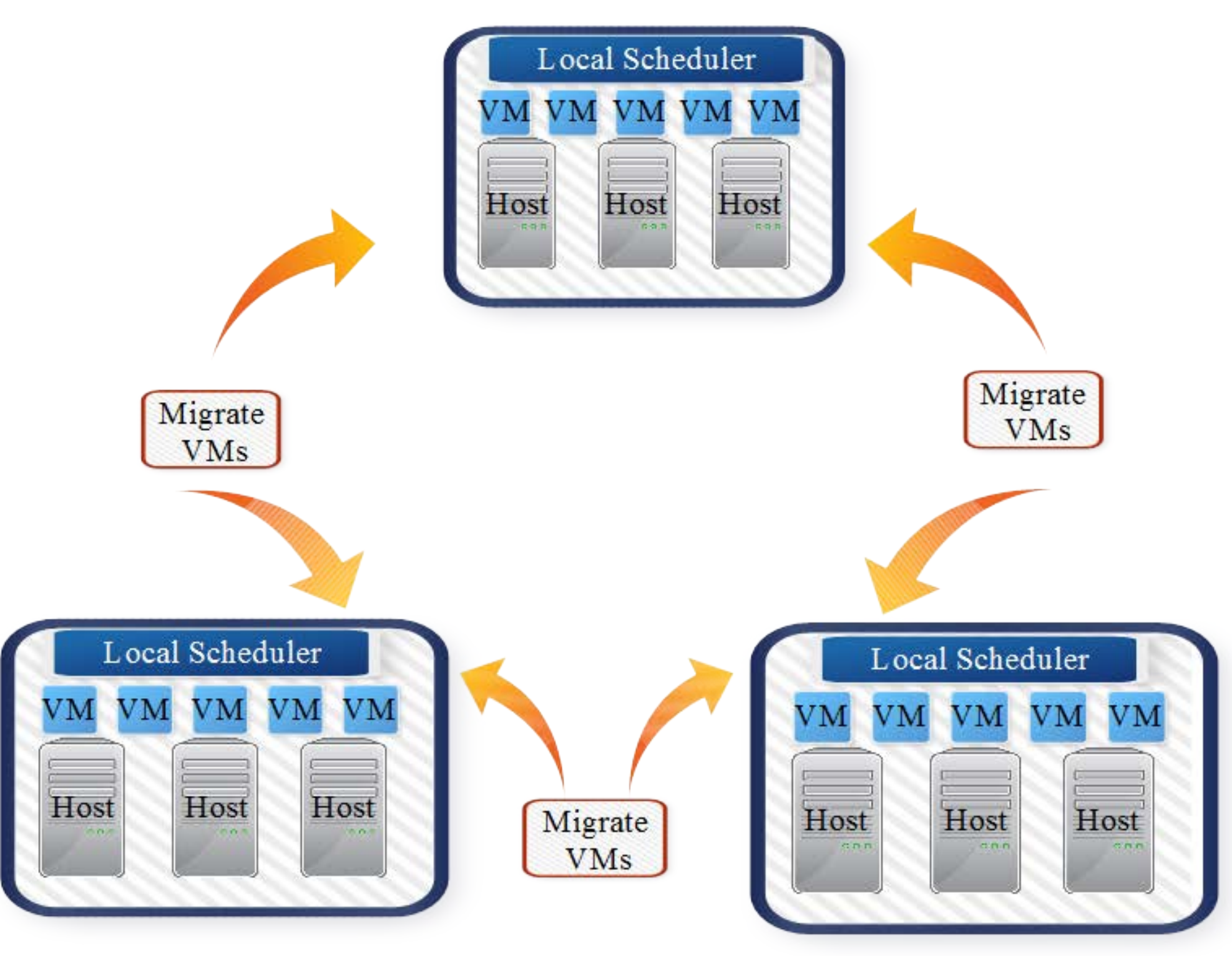}
	\caption{Distributed Scheduler}
\end{figure}

Generally, load balancing algorithms are implemented in the load schedulers, and the schedulers can be centralized or distributed.

\textbf{Centralized:} The central load balancing algorithm in Clouds are commonly supported by a centralized controller that balances VMs to hosts as shown in Fig.2, like the Red Hat Enterprise Virtualization Suite \cite{Red}. The benefits of a central management algorithm for load balancing are that it is simpler to implement, easier to manage and quicker to repair in case of a failure. Central algorithms need to obtain the global information (utilization, load, connections information, etc.), so schedulers for central algorithms are implemented as centralized to monitor information globally. The best-fit algorithm is a typical example, and other examples can also be found in \cite{Song} \cite{Ni} \cite{Tordsson} \cite{Tian} \cite{Xu}. In each execution process of the centralized algorithms, the statuses of all hosts are collected, analyzed, reordered to provide information for VM allocation. In heuristic algorithms, like greedy algorithms, the centralized scheduler allocates VMs to the hosts with the lowest load. In meta-heuristic algorithms, like genetic algorithms \cite{Thiruvenkadam} \cite{Hu}, the centralized scheduler controls crossover, mutation, interchange operations to achieve better VM-host mapping results according to fitness functions.

\textbf{Distributed:} Centralized load balancing algorithms rely on a single controller to monitor and balance loads for the whole system, which may be the system bottleneck. To relieve this problem, as shown in Fig. 3, a distributed load balancing algorithm enables the scheduling decision made by the local scheduler on each node and the associated computation overhead is distributed. 

The distributed algorithm eliminates the bottleneck pressure posed by the central algorithm scheduler and improves the reliability and scalability of the network. While the drawback of distributed algorithm is the cooperation of a set of distributed scheduler and the control plane overhead is required. This overhead should be taken into consideration when comparing the performance improvement \cite{Christ}. Cho et al. \cite{Cho} proposed ACOPS by combining ant colony optimization and particle swarm optimization together to improve VM load balancing effects and reduce overhead by enhancing convergence speed.



\section{VM Load Balancing Algorithm Modeling in Clouds }

In this section, we will discuss the details about VM load balancing algorithm design. Basically, the algorithm should consider VM model， including VM resource type, VM type uniformity, allocation dynamicity, optimization strategy and scheduling process. 

\subsection{VM Resource Type} 
When designing load balancing algorithm for VMs, the administrator can focus on single resource type or multiple resource type for scheduling. 

\textbf{Single Resource Type:} In this category, the VM resource type that is considered for balancing is limited to single resource type, generally the CPU resource. This assumption is made to simplify the load balancing process without considering other resource types,  which is common in balancing VMs running computational intensive tasks. 

\textbf{Multiple Resource Type:} 
Multiple resource type is considered in some algorithms, which monitors not only CPU load but also memory load or I/O load. These algorithms admit the fact that cloud provider offers heterogeneous or other resource-intensive types of VMs for resource provision. The general techniques to deal with multiple resource type are through configuring different resources with weights \cite{Tian}\cite{Tian2}\cite{Wen} or identifying different resources with priorities \cite{Ni}.

\subsection{VM Type Uniformity}
In VM load balancing algorithms, the VMs for scheduling are modelled as homogeneous or heterogeneous.

\textbf{Homogeneous:} In this category, VM instances offered by cloud provider are limited to a homogeneous type. Like the single resource type, this assumption is also made to simplify the scheduling process and ignores the diverse characteristic of tasks. However, this assumption is rarely adopted in a real cloud environment, because it fails to take full advantage of the heterogeneous nature of cloud resource. 

\textbf{Heterogeneous:} Cloud service providers have offered different types of VMs to support various task characteristics and scheduling objectives. For example, more than 50 types of VMs are provided by Amazon EC2, and the VMs are classified as general purpose, compute optimized and memory optimized \cite{Amazon}. In this model,  based on the task characteristic and scheduling objectives, the algorithm selects the corresponding type of hosts to allocate. 

\subsection{VM Allocation Dynamicity}
Based on VM allocation dynamicity, load balancing algorithms for VM allocation can be classified as static or dynamic: 

\textbf{Static:} Algorithms in this class are also noted as offline algorithms, in which the VMs information are required to be known in advance. Thus, static algorithms generally obtain better overall performance than dynamic algorithms. 
However, demands are changing over time in real clouds. Thus, static resource allocation algorithms are easy to violate the requirements of dynamic VM allocation. 

\textbf{Dynamic:} Algorithms in this class are also noted as online algorithms, in which VMs are dynamically allocated according to the loads at each time interval. The load information of VM is not obtained until it comes into the scheduling stage. These algorithms could dynamically configure the VM placement combining with VM migration technique. In comparison to static algorithms, dynamic algorithms have higher competitive ratio.

\subsection{Optimization Strategy}
As an NP-hard problem, it is expensive to find the optimal solutions for algorithms. Therefore, the majority of proposed algorithms are focusing on finding approximate solutions for VM load balancing problem. For this category, we classify the surveyed algorithms as three types: heuristic, meta-heuristic and hybrid. 

\textbf{Heuristic:} Heuristic is a set of constraints that aim at finding a good solution for a particular problem \cite{ Talbi}. The constraints are problem dependent and are designed for obtaining a solution in a limited time. In our surveyed algorithms, algorithms have various constraints, like number of migrations, SLAs, cost, etc., thus, the optimization functions are constructed in different ways. The advantage of heuristic algorithms is that they can find a satisfactory solution efficiently, especially in limited time cost. In addition, heuristic algorithms are easier to implement in comparison to meta-heuristic algorithms. As heuristic algorithms run fast, they are suitable for online scheduling that requires system to response in time. Greedy algorithm is a type of heuristic algorithms and is applied in \cite{Song} \cite{Ni} \cite{Tian} to quickly obtain a solution for online scheduling scenario. 

\textbf{Meta-heuristic:} Different from heuristic algorithms, meta-heuristic algorithms are mainly designed for a general purpose problem \cite{ Talbi}. Therefore, meta-heuristic algorithms follow a set of uniform procedures to construct and solve problems. The typical meta-heuristic algorithms are inspired from nature, like Genetic algorithms, Ant Colony Optimization, Particle Swarm Optimization and Honeybee Foraging algorithms. These algorithms are based on population evolutions and obtaining the best population in each evolution and keep it into next evolution. A distributed VM migration strategy based on Ant Colony Optimization is proposed in \cite{Wen}. Ant Colony Optimization and Particle Swarm Optimization are combined in \cite{Cho} to deal with VM load balancing. The results in these proposed strategies show that better load balancing effects can be achieved compared to heuristic algorithms. However, in comparison to heuristic algorithms, meta-heuristic algorithms need more time to run and find the final solution as its solution space can be quite large. Moreover, the meta-heuristic are generally stochastic processes and their convergence time and solution results depend on the nature of problem, initial configurations and the way to search the solutions. 
	
\textbf{Hybrid:} For hybrid algorithm, heuristic algorithm is used to fulfill the initial VM placement and meta-heuristic algorithm is used to optimize the placement of VMs during migration. Alternatively, meta-heuristic algorithms can be applied firstly to generate a set of solutions, and then heuristic algorithms are used to obtain the optimized solution based on these solutions. In either way, the time cost and solution space are both reduced, while the implementation complexity increases. Thiruvenkadam et al. \cite{Rouzaud} proposed a hybrid genetic algorithm that follows the first approach. 

\subsection{Scheduling Process Modeling}
The load balancing scheduling process can be mainly divided into VM initial placement stage and VM live migration stage. 

Some research has focused on the VM load balancing at the initial placement stage without considering live migration \cite{Xu} \cite{Ni} \cite{Tordsson} \cite{Wen}\cite{YangXing}. At this stage, the key component of the scheduling process is the VM acceptance policy, which decides the host placement that the VM is allocated to. The policy generally takes the host available resource into consideration.

As for the live migration stage in scheduling process, it mainly considers following aspects: 

(1) VM migration policies enable cloud data centers to establish preferences when VMs are migrated to other hosts. The VM migration policies indicate when to trigger a VM migration from one host to another host. Generally, they consist of a migration threshold to trigger migration operations, and the threshold is decided by a data center administrator based on the computing capabilities of each host, such as in Red Hat \cite{Red} and VMware \cite{VMware}. For instance, a CPU-intensive host may be configured with a relatively high threshold on CPU usage, while an I/O intensive host may be configured with a relatively low threshold on CPU usage.

(2) VM selection policies enable cloud data centers to establish polices to select which VMs should be migrated from overloaded hosts. Generally, an overloaded host has a high probability to host too many VMs. The VM selection policies firstly need to select the overloaded hosts.
The VM selection policies also decide which VMs should be migrated to reduce the load of the overloaded host as well as satisfy other objectives, like minimizing the number of migrations \cite{Hu} \cite{Cho} and reducing migration latency \cite{Hu}. 

(3) VM acceptance policies enable cloud data center to establish approaches about which VMs should be accepted from other overloaded hosts in the process of balancing loads collaboratively among hosts via VM live migration. 
The VM acceptance policies need to collect information, such as a) remaining resource of hosts, b) an associated resource type either CPU or memory, and c) a threshold either above or below a certain remaining resource amount. 
Then, the VM acceptance policies are applied to determine whether to host a given VM.

\section{Load Balancing Scheduling Metrics Comparison}
For VM load balancing, there are different metrics to evaluate the performance of load balancing algorithms. These metrics are optimized based on different behaviors, like obtaining maximal or minimal values. In this section, we introduce prominent metrics adopted in VM load balancing algorithms, like utilization standard deviation, makespan, etc. Table I lists the metrics adopted in our surveyed algorithms and their optimization behavior.

\textbf{Load Variance and Standard Deviation of Utilization:} Both of these two metrics specify the deviation from the mean utilization. These metrics are quite popular in some articles as they are easy to be measured. However, for some other load balancing algorithms focusing more on time constraint rather than utilization, they are not appropriate.

\textbf{Makespan:} Makespan is the longest processing time on all hosts and it is one of the most common criteria for evaluating a scheduling algorithm. Sometimes, keeping the load balanced is to shorten the makespan, and a shorter makespan is the primary purpose of a scheduling algorithm \cite{Cho}. Compared with metrics like load variance or standard deviation of utilization, it pays more attention to time constraint, which is better for evaluating real-time scheduling load balancing algorithms.


\textbf{Number of Overloaded Hosts:} It measures how many hosts in Clouds are overloaded, which gives an overview of the system status. And this value is dependent on the pre-configuration of overloaded threshold. Load balancing algorithms aim to reduce the number of overloaded hosts as much as possible. This is a straightforward metric to evaluate load balancing effect, but it gives few details about loads distribution.


\textbf{Percent of all VMs to be Located:}  It is applied to VM load balancing in multiple data centers and specifies the VM distribution percentage of different data centers as constraints. Its values are established with a minimum and maximum percentage of all VMs that can be located in each Cloud. Combining the these values and applying the integer programming formulation, the numbers of VMs allocated in multiple Clouds are balanced \cite{Tordsson}. However, since the balance is only based on the number of VMs and does not consider VM resource amount, if the VMs are heterogeneous, the VM load balancing effects are still open to be discussed. 


\textbf{Quadratic Equilibrium Entropy:}  It is motivated  by the situation that ideal load balancing algorithms maintain load equilibrium during the scheduling time period, and the information entropy measures the average equilibrium uncertainty  \cite{Wang}. The information entropy  is based on the theory of linear equilibrium entropy and quadratic equilibrium entropy. With greater entropy, more balanced loads are distributed. This metric offers a new option to evaluate performance of different load balancing algorithms. 


\textbf{Throughput:}  It measures how fast the hosts can handle with requests, as imbalanced loads may reduce system performance. Therefore, higher throughput comes along better system load balancing  situation. It is suitable for scenarios that care about service response time. For load balancing algorithms, generally, this metric is not evaluated individually,  and it is often evaluated with other metrics, like in \cite{Rouzaud}, number of migrations is measured together with throughput.

\textbf{Standard Deviation of Connections:}  It is regarded as a kind of loads in \cite{Bhadani} that focuses on the connections. To some degree, its meaning is similar to the standard deviation of utilization. This metric suits for the network-intensive systems. However, different connections may consume different amount of resource, this metric does not represent the resource usage.

\textbf{Average Imbalance Level:} The popular metric like the standard deviation of utilization only considers a single type of resource, like CPU utilization. The average imbalance level metric considers multiple types of resource together, like CPU, memory and bandwidth together. It measures the deviation of these resource respectively on all the hosts and then combines them together with weights to denote the load balance effects \cite{Tian}. This metric is available for the scenario that multiple resource may be the bottleneck, but service providers need efforts to identify the appropriate weights for their resource.

\textbf{Capacity-makespan:} It combines the load and requests lifecycle together compared with traditional metrics without considering lifecycle. It is derived from the makespan metric \cite{Tian2}. Traditionally, the makespan is the total length of processing time, while capacity-makespan is defined as the sum of the product of required capacity (resource) and its processing time. This metric reflects the feature of capacity sharing and fixed interval constraint in Clouds, and it is more suitable for Clouds with reservation model. In reservation model, resources are allocated to requests with fixed amount of resources or time intervals.


\textbf{Imbalance Score:} It represents the degree of overload of a host based on exponential weighting function, which aims to overcome the limitation of linear scoring \cite{Singh}. This metric provides reference about how high the host utilization is above the predefined threshold and also considers the multiple resource. The system total imbalance score is computed as the sum of all hosts imbalance score. Therefore, the load balancing algorithms target to minimize this metric if they adopt it.

\textbf{Remaining Resource Standard Deviation:} It measures the standard deviation of available resource of hosts that can be allocated to VMs \cite{Thiruvenkadam}. The standard deviation of utilization is measured with the used resource, while this metric measures  the remaining resource. The disadvantage of this metric also lies in that it is not suitable for algorithm that focuses on time constraint.

\textbf{Number of Migrations:} This is an auxiliary metric that represents the performance and is measured with other metrics together. Too many migrations may achieve balanced loads, but lead to performance degradation, therefore, it is a tradeoff metric between load balancing and performance. It is not reasonable to use this single metric to evaluate load balancing effects. 

\textbf{SLA Violations:} This is another auxiliary metric that represents the performance. SLA violation can be defined as a VM cannot fetch enough resources  (like CPU mips \cite{Wen}) from host. Too many SLA violations show that the hosts are not balanced well, thus this metric should be minimized. Since it is also an auxiliary metric, like the number of migrations, this metric should be evaluated together with other metrics.

\begin{table}
\caption{Metrics in Our Surveyed Paper} 
\newcommand{\tabincell}[2]{\begin{tabular}{@{}#1@{}}#2\end{tabular}}
\centering 
\scriptsize
\begin{tabular}{|l|l|l|} 
\hline\hline 
\textbf{Metrics} & \textbf{Optimization Behavior} & \textbf{\tabincell{c}{Algorithm}} \\
\hline 
Load Variance and Standard Deviation of Utilization & Minimize & \cite{Yang} \cite{Ni} \cite{Zhao} \cite{Zhou}\\
\hline
　Makespan & Minimize &  \cite{Cho}\\
\hline
Number of Overloaded Hosts &Minimize &  \cite{Song}\\
\hline
Percent of all VMs to be Located in Host & Minimize and Maximize &  \cite{Tordsson}\\
\hline
Quadratic Equilibrium Entropy & Minimize &  \cite{Wang} \cite{Yang}\\
\hline
Throughput & Improve & \cite{Rouzaud} \cite{Bhadani}\\
\hline
Standard Deviation of Connections & Minimize &  \cite{Bhadani}\\
\hline 
Average Imbalance Level & Minimize &\cite{Tian}\\
\hline 
Capacity-makespan & Minimize & \cite{Tian2} \cite{Xu}\\
\hline 
Imbalance Score & Minimize & \cite{Singh}\\
\hline 
Remaining Resource Standard Deviation & Minimize & \cite{Thiruvenkadam}\\
\hline 
Number of Migrations & Reduce or Minimize & \cite{Hu} \cite{Wen}\\
\hline 
SLA Violations & Minimize & \cite{Wen}\\


\hline 

\hline 
\end{tabular}
\label{tab:LPer}
\end{table}

\section{Performance Evaluation Approaches }
In this section, we will discuss some realistic platforms and simulation toolkits that have been adopted for VM load balancing performance evaluation as illustrated in Fig. 4.
\subsection{Realistic Platforms}
Conducting experiments under realistic environment is more persuasive and there exist some realistic platforms for performance testing.

\textbf{OpenNebula:} It is an open source platform that aims at building industry standard open source cloud computing tool to manage the complexity and heterogeneity of large and distributed infrastructures. It  also offers rich features, flexible ways, and better interoperability to build clouds. By combining virtual platforms, like KVM, OpenNebula Cloud APIs for VMs operations and Ganymed SSH-2 for resource information collection, new VM load balancing algorithm could be implemented and tested \cite{OpenNebula}.

\textbf{ElasticHosts:} It is a global cloud service provider containing geographical diverse distributions that offer easy-to-use Cloud Servers with instant, flexible computing capacity. Apart from Cloud Servers, Elastichosts also offers Managed Cloud Servers, Cloud Websites, and Reseller Programs, which are easy for developers to do research \cite{ElasticHosts}.

\textbf{EC2}: Amazon EC2 is a commercial Web service platform that enables customers to rent computing resources from the EC2 cloud. Storage, processing and Web services are offered to customers. EC2 is a virtual computing environment, which enables customers to use Web service interfaces to launch different instance types with a variety of operating systems \cite{Amazon}.

There are some other popular cloud platforms, like Eucalyptus, CloudStack and OpenStack, while they are not applied to evaluate VM load balancing in our surveyed papers, thus, we do not introduce them in detail.

\subsection{Simulation Toolkits}
Concerning unpredicted network environment and laboratory resource scale (like hosts), sometimes it is more convenient for developing and running simulation tools to simulate large scale experiments. The research on dynamic and large-scale distributed environment can be fulfilled by constructing data center simulation system, which offers visualized modeling and simulation for large-scale applications in cloud infrastructure \cite{Tian4}. The data center simulation system can describe the application workload statement, which includes user information, data center position, the amount of users and data centers, and the amount of resources in each data center \cite{Tian3}. Under the simulated data centers, load balancing algorithms can be easily implemented and evaluated. 

\textbf{CloudSim:} CloudSim is an event driven simulator implemented in Java. Because of its object-oriented programming feature, CloudSim allows extensions and definition of policies in all the components of the software stack, thereby making it a suitable research tool that can mimic the complexities arising from the environments \cite{Calheiros}.

\textbf{CloudSched:} CloudSched enables users to compare different resource scheduling algorithms in IaaS regarding both hosts and workloads. It can also help the developer identify and explore appropriate solutions considering different resource scheduling algorithms \cite{Tian4}. 

\textbf{FlexCloud:} FlexCloud is a flexible and scalable simulator that enables user to simulate the process of initializing cloud data centers, allocating virtual machine requests and providing performance evaluation for various scheduling algorithms \cite{Xu2}.  \\

\begin{figure}
	\centering
	\includegraphics[width=0.9\textwidth, angle=-0] {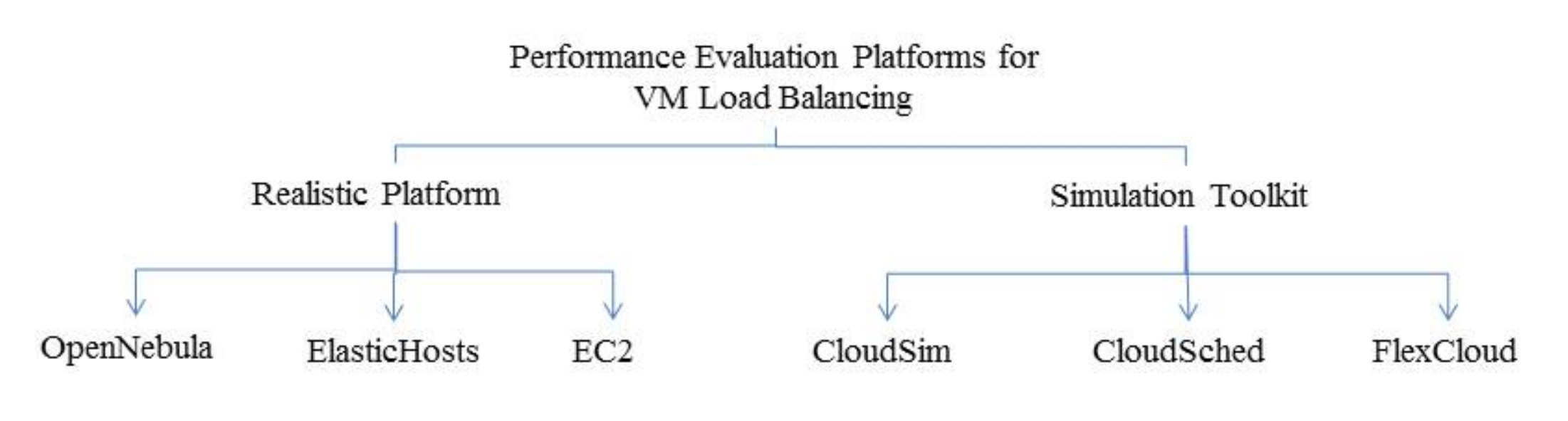}
	\caption{Performance Evaluation Platforms for VM Load Balancing}
\end{figure}


 Table II summarizes approaches used by authors to evaluate their VM load balancing algorithms. We also list their experimental scenarios and performance improvement achieved by them.
The experimental environment contains the information about the experimental platforms and scale. Under realistic platforms, the number of machines for testing is almost less 10,  but in simulations, the hosts and VMs scale are increased to hundreds and thousands.
The performance improvements include the percentage of load balancing effect improvements based on different metrics. The performance also shows that some algorithms significantly improve the VM load balancing effect. Some of our surveyed papers compare their algorithm with the same baselines, like \cite{Tian}\cite{Tian2}\cite{Thiruvenkadam} all select round-robin algorithm as one of their baselines. While these algorithms are rarely compared with each other, which leads to a future work that we will discuss in Section 7.




\begin{table}[h]
	\caption{A Summary of Environment Configuration and Performance Improvement of VM Load Balancing Algorithms Noted by Respective Papers} 
	\newcommand{\tabincell}[2]{\begin{tabular}{@{}#1@{}}#2\end{tabular}}
	\centering 
	\scriptsize
	\begin{tabular}{|c|c|c|} 
		\hline\hline 
		\textbf{Algorithm} & \textbf{Experiments Configuration} & \textbf{\tabincell{c}{Performance Improvement}}\\
		\hline
		Song et al. \cite{Song} & \tabincell{c}{10 heterogeneous hosts with CentOS \\ and Xen hypervisor} & \tabincell{c}{It saves 22.25\% average execution time compared with \\static distribution algorithm when reaching same load\\ balancing level. } \\ 
		\hline
		Ni et al. \cite{Ni} & \tabincell{c}{Based on OpenNebula, virtual platform \\is KVM, hosts are 6 IBM BladeCenter \\Servers, both CPU resource and memory \\resource are considered} & \tabincell{c}{When VMs loads increase, it reduces more \\imbalance effects for any type of resource compared \\with the single type of resource in OpenNebula.} \\
		\hline
		Tordsson et al. \cite{Tordsson} & \tabincell{c}{ElasticHosts and EC2 cloud with two data \\centers (in the USA and in Europe), \\containing 4 types of instances} & \tabincell{c}{Through configuring the minimum percent of VMs \\to be placed in each cloud under multi-cloud \\environment to balance load, it could save more \\budget than single cloud.} \\
		\hline
		Zhao et al. \cite{Zhao} & \tabincell{c}{4 hosts with OpenVZ for managing VMs} & \tabincell{c}{The algorithm convergences fast and keeps the standard \\deviation of load in a low range.} \\
		\hline
		Yang et al.\cite{Yang} & \tabincell{c}{Simulation with 20 hosts} & \tabincell{c}{Compared with no load balancing and minimum \\connection algorithm, it  reduces the number \\of overloaded hosts.} \\
		\hline
		Bhadani et al. \cite{Bhadani} & \tabincell{c}{Hosts installed with CentOS and Xen \\kernel, as well as Apache web server} & \tabincell{c}{Tests are conducted on limited capacity and results \\show that the algorithm  improves up to 20\% \\throughput has better load balancing effects \\compared with isolated system.} \\
		\hline
		Rouzaud-Cornabas \cite{Rouzaud} & \tabincell{c}{Simulation with more than 100 \\heterogeneous hosts and 1000 \\heterogeneous VMs } & \tabincell{c}{About 10\% faster to detect overloaded hosts and \\solve the overloaded situation to reach predefined \\balanced situation, compared with algorithm \\without its load balancing mechanism.} \\
		\hline
		Tian et al. \cite{Tian} & \tabincell{c}{Simulation under CloudSched with \\hundreds of heterogeneous hosts and \\thousands heterogeneous  of VMs} & \tabincell{c}{It reduces 20\%-50\% average imbalance value \\\ compared with its baselines} \\
		\hline
		Tian and Xu \cite{Tian2} & \tabincell{c}{Simulation under CloudSched with \\hundreds of heterogeneous hosts and \\thousands heterogeneous  of VMs} & \tabincell{c}{It has 8\%-50\% lower average makespan and \\\ capacity-makespan than its baselines, such as \\\ Longest Processing Time first and RR algorithms} \\
		\hline
		\tabincell{c}{Thiruvenkadam \\et al. \cite{Thiruvenkadam}} & \tabincell{c}{Simulation with CloudSim} & \tabincell{c}{It has lower load imbalance value compared with RR, \\\ First Fit, and Best Fit algorithms.} \\
		
		\hline
		Hu et al. \cite{Hu} & \tabincell{c}{6 hosts based on OpenNebula, virtual \\platform is KVM; hosts are connected \\with LAN} & \tabincell{c} {When the system load variation is evident, it  guarantees \\ the system load balancing better compared with \\ Least-loaded scheduling algorithm and rotating \\ scheduling algorithm.} \\
		
		\hline
		Wen et al. \cite{Wen} & \tabincell{c}{Simulation with CloudSim with 2 types \\of hosts and 4 types of VMs under \\random workload} &  \tabincell{c}{ It reduces about 40\%-70\% load variance compared \\ with the baselines offered in CloudSim.}   \\
		
		\hline
		Cho et al. \cite{Cho} & \tabincell{c}{Simulation on a personal computer} & \tabincell{c}{ It reduces 5\%-50\% makespan, compared with\\ other genetic algorithms, and no worse than First \\ Come First Serve + RR algorithms.} \\
		\hline

		\hline 
		
		\hline 
	\end{tabular}
	\label{tab:LPer}
\end{table}

\section{Algorithms Comparison}
In this section, we will discuss a few VM load balancing algorithms with the classifications discussed in the previous section. 
\subsection{Migration Management Agent}
Song et al. \cite{Song}  proposed a Migration Management Agent (MMA) algorithm for dynamically balancing virtual machine loads in High Level Application (HLA) federations. 
For HLA systems, especially large-scale military HLA systems, their computation and communication loads vary dynamically during their execution time. In this algorithm, VMs are allowed to be migrated between different federations to balance the loads while the communication costs are also incurred. Therefore, the objectives of this algorithm are twofold: reducing the load of the overloaded hosts and decreasing the communication costs among different federations. 
Prior to introduce their VM load balancing algorithms, the authors predefined host utilization threshold for detecting overloads, and modelled host and VMs loads based on CPU utilization. They also modelled communication costs for VMs on the same host and different hosts respectively, as the communication costs in a local host consumes much less communication resource than among different hosts. The MMA algorithm applies live migration to migrate VMs from overloaded hosts to the least loaded host and ensures the migration would not make the destination hosts overloaded. As a heuristic, the algorithm also calculates the communication costs between VMs and hosts, and selects the migration path with the least communication costs. From the results based on both realistic platform and simulation, it is observed that the number of overloaded hosts is reduced.
%

The advantage of MMA is that it considers and models communication costs between the migrated and the rest VMs, and it could dynamically balance loads under communication constraints. While its disadvantage is that it neglects the stochastic interaction characteristics between VMs and hosts. Apart from that, only CPU utilization is considered to be the load of hosts.

\subsection{Virtual Machine Initial Mapping based on Multi-resource Load Balancing}
Ni et al. \cite{Ni}  presented a VM mapping algorithm considering multiple resources  and aimed at easing load crowding, which is based on the probability approach to adapt unbalanced loads. The authors focused on the scenario with concurrent users. The concurrent users may simultaneously require the same resource from the same host, increasing the loads of target host rapidly and leading the performance degradation.
Multiple resources are considered with weights in the proposed algorithm. With the weighted resource, each host has its corresponding score that is inverse proportional to its utilization. The algorithm also uses Proportional Selection to compute the selection probability of each host, in which the host with the higher score has the higher probability to accept VMs. Although this approach is based on probability calculation, it is a deterministic approach rather than stochastic one, as both the hosts utilization and their scores are determined. Therefore, this approach still belongs to heuristic strategy.


The realistic experiment based on homogeneous VMs shows that this approach could efficiently reduce the standard deviation of utilization of all nodes, while this algorithm mainly focuses on the initial placement of VMs rather than in the running stage.

\subsection{Scheme for optimizing virtual machines in multi-cloud environment}
The algorithm proposed by Tordsson et al. \cite{Tordsson} for VM placement optimization aims to multi-objective schedule including load balancing, performance and cost. 
As in a multi-cloud, different cloud providers are supported by different infrastructures and offer different VM types, the authors spent their efforts on handling with heterogeneous resource under multi-cloud. 
The proposed algorithms are embedded in a cloud broker, which is responsible for optimizing VMs placement and managing the multiple virtual resource. The authors explore a set of meta-heuristic algorithms that are based on integer programming formulations and their formulation is a version of Generalized Assignment Problem. These algorithms mainly focus on performance optimization, like makespan, throughput and network bandwidth usage. The intensive experiment results show that multi-cloud placement can reduce costs under load balancing constraints.

This work has comprehensive experiments and comparisons while it mainly considers the static scheduling for VMs rather than dynamic. Therefore, the  scalability of algorithms would be limited when they are applied to the dynamic scenario.

\subsection{DLBA-CAB}
To balance intra-cloud, Zhao et al. \cite{Zhao} presented a distributed load balancing algorithm based on comparison and balance (DLBA-CAB) by adaptive live migration of virtual machines. 
The algorithm was initially designed to enhance EUCALYPTUS \cite{Nurmi} by complementing load balancing mechanism.
Its objective is making each host to achieve equilibrium of processor usage and I/O usage. The authors modelled a cost function considering weighted CPU usage and I/O usage, and each host calculates the function values individually. In each monitor interval, two hosts are selected randomly to build a connection to find the cost difference between them. The difference is regarded as migration probability, in which the VMs are always migrated from the physical hosts with a higher cost to those with a lower one. During the live migration, the algorithm also aims to minimize the host downtime to improve the system stability. After migration, the algorithm enables the system to reach a Nash equilibrium that reflects the loads are well balanced. This algorithm does not need a central coordinator node while the loads information of other hosts would be stored on shared storage and updated periodically. The realistic experiments have shown that this heuristic keeps the deviation of loads in a low level.

DLBA-CAB is an example showing how distributed load balancing algorithm for VMs is implemented in intra-cloud with fast convergence speed to reach Nash equilibrium while its model simply assumes that host memory usage is always enough. 

\subsection{Optimized Control Strategy Combining Multi-Strategy and Prediction Mechanism}
Yang et al. \cite{Yang} designed a multi-strategy based on prediction mechanism to reduce the number of overloaded hosts and avoid unnecessary migration. 
The authors also adopted a weighted function considering multiple types of resource, as same as the algorithms introduced in Section 6.2 and 6.4. 
To identify the load of hosts, they defined four status domains: light-load, optimal, warning and overload, for different utilization domains. Hosts with different utilization lie in different domains, and different migration strategies are executed in different domains. Moreover, to analyze and predict future utilization for resource components, this strategy contains a prediction model that uses a set of recently utilization data series based on an AR prediction model \cite{Gersch} to obtain the future utilization.  
As for choosing the migration destination placement, this strategy considers the characteristic of applications, like CPU intensive and I/O intensive. The migration destination is selected as the host that is most suitable for the predicted resource change, like if the CPU fluctuation trend is the most influential one, the host with the largest CPU resource is selected as the destination.  
In addition to the migration process, to avoid the multiple VMs migrating to the same host and overloading simultaneously, a three times handshaking protocol is used to confirm the ultimate migration. With this protocol, each host maintains an acceptance queue containing VMs that are waited to be allocated, and this queue updates host utilization load increment along with time. The simulation results prove that this heuristic is efficient to reduce the number of overloaded hosts and migration time.   

The advantage of this algorithm is its adaptivity that different strategies are applied to different host status respectively, which ensures the algorithm to be adaptive to various situations. However, this algorithm is only evaluated with small scale hosts and not tested under realistic platforms.

\subsection{Central Load Balancing Policy for VM}
Bhadani et al. \cite{Bhadani} proposed a Central Load Balancing Policy for VM (CLBVM) to balance loads evenly in Clouds. The authors designed this policy for distributed environment to achieve shorter response time and higher throughput. To achieve these goals, the policy requires several characteristics: 1) low overhead is generated by load balancing algorithm, 2) load information is updated and collected periodically, 3) minimum downtime is caused by live migration.  This policy is based on global state information and the migration operation is a mix of both distributed and centralized. In this heuristic, on each host, the load information collector collects its CPU load information continuously (hosts would be labeled as Heavy, Moderate, and Light based on different load levels) and exchanges information with a master server, which periodically reallocates the loads on heavily loaded host to the lightly loaded host.

This policy advances the existing model for load balancing of VM in a distributed environment and the practice in XEN shows its feasibility to improve throughput. While this policy simply assumes that the network loads are almost constant, this is not very applicable to current Cloud environment. In addition, another limitation is that the resource type of memory and I/O are rarely considered in this work.

\subsection{Distributed Dynamic Load Balancer for Virtual Machine }
Rouzaud-Cornabas \cite{Rouzaud} presented a distributed dynamic load balancer for VMs based on P2P architecture. Its objectives are reducing the load on a single host, moving a VM to a new host with more resources or with specialized resources. The author chose dynamic scheduling since the VM behaviors cannot be precisely predicted because of complex behaviors and non-deterministic events. The author also aimed to achieve better system scalability, therefore, the load balancers are designed as distributed ones to overcome the scalability bottleneck of the single load balancer.
To balance the loads, the author adopted a score function composite of the static score and dynamic score to represent the loads. The static score takes into account static resource quota reserved for a VM, and the dynamic score mainly considers the dynamic resources like the amount of free memory. 
After calculating the scores on all hosts, in the placement and migration processes, the algorithm selects the host that fits the static requirement of VMs to be their destination. The simulation results demonstrate that the proposed approach speeds up the time to detect and solve overloaded situation.

In this approach, the load balancer on each host cooperates together to ensure the system scalability and does not need centralized control. However, communication cost may increase rapidly when the number of hosts becomes more, which is not considered in this article.

\subsection{Dynamic and Integrated Resource Scheduling Algorithm}
Tian et al. \cite{Tian} introduced a dynamic and integrated resource scheduling algorithm (DAIRS) for balancing VMs in Clouds. This algorithm treats CPU, memory and network bandwidth as integrated resource with weights.  They also developed a new metric, average imbalance level of all the hosts (details are given in Section 4), to evaluate the performance under multiple resource scheduling.  
In DAIRS, VM requests are processed as like in a pipeline. VM requests are identified at different statuses and put into different queues to process. For example, VMs that are waiting for allocation are put into the waiting queues, and VMs that need reallocation are put into the optimization queue to be migrated. If the VM status is changed, the VM is transferred to another queue and processed. Thus, the VMs management is converted to queue management. The algorithm monitors system load information at each time interval and VMs are allowed to be delayed allocation if the host during a time interval is overloaded. If overloading occurs, the VMs on the overloaded hosts (also in the optimization queue) are migrated to the host with the least load.
The simulations conducted with heterogeneous hosts and VMs showed that DAIRS have reduced 20\% to 50\% average imbalance level than baselines.

DAIRS is one of the earliest algorithms that explored the multiple types of resources and treated them as integrated value. The main drawback of DAIRS is that it ignores the communication cost of migrations. 

\subsection{Prepartition}
Tian and Xu \cite{Tian2} designed an algorithm for offline VM allocation within the reservation model, namely Prepartition. As VMs requests are reserved, all VM information has been known before the final placement. Thus, in the reservation model, the VMs requests are partitioned into smaller ones to utilize resource better and reduce overloads. VMs with multiple resource are considered in this paper. The authors also redefined the traditional metric makespan as a new metric capacity-makespan, which is computed as VM CPU load multiplies VM capacity. The VM requests are partitioned with a partition value that is 
calculated as the larger value between the average capacity-makespan and maximum capacity-makespan of all VMs. A partition ratio (a positive integer) that represents how many parts are desired to be partitioned is also defined by the authors. Then, each VM is partitioned into multiple VMs with the length equivalent to  the partition value divided by partition ratio. After the VMs with smaller size are generated, the VMs are allocated one by one to the host with the lowest capacity-makespan. It is noticed that the regeneration process is before the final placement, therefore, it may not cause the instability and chaos. Simulated with heterogeneous cloud and real traces, the authors illustrated that Prepartition algorithm achieved lower average makespan and capacity-makespan than baselines.

Though belonging to the static algorithm, Prepartition is efficient to achieve better load balance as desired. For offline load balancing without migration, the best approach has the approximation ratio as 4/3 \cite{Graham}. With approximation ratio analysis, the authors have proved that the approximation ratio of Prepartition is possible to be approaching the optimal solution.

\subsection{Hybrid Genetic based Host Load Aware Algorithm}
Thiruvenkadam et al.  \cite{Thiruvenkadam} presented a hybrid genetic algorithm for scheduling and optimizing VMs. One of their objectives is minimizing the number of migrations when balancing the VMs. The authors paid more attention to the variable loads of hosts and dynamicity of VM allocations. Therefore, the algorithm considers two different techniques to fulfill these goals: one is that initial VM packing is done by checking the loads of hosts and user constraints, and the other is optimizing VMs placement by using a hybrid genetic algorithm based on fitness functions. Furthermore, a centralized controller is needed to store hosts historical and current loads globally. Similar to \cite{Tian2} described in Section 6.9, the VM optimization problem is also modelled as a bin packing problem, and both of them extend the traditional bin packing problem to be multiple dimensions by investigating multiple resource.

For the initial VMs packing, the authors proposed a heuristic approach based on multiple policies. This heuristic approach searches hosts according to VM resource requirement and host available resource to improve resource usage, which belongs to greedy algorithm. 
For the hybrid genetic algorithm for VM placement optimization, it iteratively uses different operations to generate optimized solutions. The optimization goal follows a fitness function that aims to minimize the standard deviation of the remaining resource on each host. The genetic algorithm keeps running and searching optimized solutions until the metrics are satisfied. Thus, to achieve better performance, this meta-heuristic requires more time than heuristic algorithms, such as \cite{Ni} in Section 6.2 and \cite{Zhao} in Section 6.4. Apart from the number of migrations minimization, this work investigates more optimization objects, like the number of active hosts, energy consumption and resource utilization. The simulations under CloudSim also demonstrated the trade-offs between execution time and number of migrations as well as the standard deviation of loads.

 
This approach coordinates heuristic and meta-heuristic algorithms together to achieve scheduling objectives, while this also increases the implementation complexity in realistic environment.

\subsection{VM Scheduling Strategy based on Genetic Algorithm}
Another meta-heuristic based on genetic algorithm is presented by Hu et al. \cite{Hu}, which sets its objectives as finding the best mapping solutions to achieving the best load balancing effects and minimizing migration times. As same as \cite{Thiruvenkadam} that is described in Section 6.10, the authors in this paper also addressed the load variation and used historical data to analyze. The difference is that \cite{Thiruvenkadam} applies binary codes to denote solutions, but this algorithm chooses the spanning tree structure to generate solutions. The spanning tree follows the principle that it satisfies predefined load conditions or generates relative better descendants as solutions. 
The least loaded node is set as the leaf node and has the highest probability to accept VMs. And the node with more loads are moved closer to the root node.
In the initialization stage, the authors firstly compute the selection probability of every VM, which is computed as its load divided by the sum of all VMs loads. To follow the fitness function, tree nodes are operated to optimize the placement of VMs and regenerated new trees. Each new tree represents a new solution. The algorithm repeats iteratively until it finishes the predefined loops or convergences. This approach requires a centralized controller to collect nodes (hosts) information.


This algorithm considers both the historical data and current data when computing the probabilities, which captures the influence in advance. Therefore, the algorithm is able to choose the solution that has least influence on the system after reallocation. Realistic experiments show that better load balancing performance is obtained compared with the least-loaded scheduling algorithm. However, the algorithm complexity is still open to discuss.

\subsection{Distributed VM Migration Strategy based on Ant Colony Optimization}
Wen et al. \cite{Wen} introduced a distributed VM migration strategy based on Ant Colony Optimization (ACO). The objectives of this meta-heuristic are achieving load balancing and reasonable resource utilization as well as minimizing the number of migrations. Compared with traditional centralized migration strategy, in this paper, the distributed local migration agents are able to improve system scalability and reliability. They autonomously monitor the resource utilization of each host and overcome the shortcomings of simpler trigger strategy and misuse of pheromone (the information that ants leave when they are traversing) from other ACO approaches. 
The authors redefined the pheromones as positive and negative to mark the Positive Traversing Strategy and Negative Traversing Strategy. 
The Positive Traversing Strategy represents the path that ants leave more pheromones and the Negative Traversing Strategy represents the path that ants leave less pheromones.
When overloading occurs, the distributed migration agent on each host sorts all the VMs according to their average loads. The VMs with higher load are prone to be migrated. The VMs are continued being put into a migration list until the host is not overloaded. 
The distributed migration agents are also responsible for generating some ants to traverse for new solutions. 
The ants produce more pheromones when the load on the destination host is higher or the bandwidth resource is less (through Positive Traversing Strategy). With more iterations, the ants are more likely to traverse through those hosts that are in high load condition. Finally, a list of hosts with low load condition is obtained (through Negative Traversing Strategy) and they can be matched with the sorted VMs that are prepared to be migrated, which is the final solution of the scheduling problem. 

The simulations under CloudSim toolkit with heterogeneous VMs shows that this ACO-based strategy reaches a balanced performance among multiple objectives, including the number of SLA violations, the number of migrations and load variable. However, considering the computation and time cost, VMs are scheduled in a static way that all VMs information are known in advance.

\subsection{Ant Colony Optimization and Particle Swarm Optimization}
Cho et al. \cite{Cho} combined ant colony optimization and particle swarm optimization (ACOPS) to deal with VM load balancing in Clouds. Its objectives are maximizing the balance of resource utilization and accepting as many requests as possible. 
Compared with other meta-heuristics that schedule VMs in a static way, like \cite{Tordsson} introduced in Section 6.3 and \cite{Wen} introduced in Section 6.12, this meta-heuristic optimizes VM placement in a dynamic way. 
The authors considered both CPU and memory resource to schedule. To reduce solution dimensions and execution time, this algorithm adopts an accelerating step, namely Pre-reject, in which the remaining memory of each server is checked  before scheduling. 
If the maximum remaining memory is less than the memory demand of a request, the VM request is rejected. 
To construct an initial solution from all the ants, the authors predefined the probability for ants to search the next path.  
The algorithm then applies Particle Swarm Optimization (PSO) to improve the results by using the global best solution to generate a better solution. In each iteration, a fitness function is applied to evaluate the performance from all the solutions finished completely by all the ants. 
Instead of using both global and local pheromone update that cost a large amount of time, the algorithm only applies global pheromone update so that the paths belonging to the best solution may occupy increased pheromone. Finally, ACOPS is terminated when the iteration reaches predefined iterations or the global best solution keeps constant during a given time, just like other meta-heuristics.

As a complementary for other ACO and PSO algorithms, the time complexity of ACOPS is induced by the authors. In addition, the results demonstrate the algorithm effectiveness in balancing loads. Although the Pre-reject step accelerates the process to obtain a solution, it also rejects a set of VMs, which leads to revenue loss of Cloud service providers.

\begin{table}[]
	\centering
	\caption{Algorithm Classification for VM Model}
	\scriptsize
	\label{my-label}
	\begin{tabular}{|c|c|c|c|c|}
		\hline
		\textbf{Algorithm} &  \textbf{\begin{tabular}[c]{@{}l@{}}VM Allocation \\ Dynamicity\end{tabular}} & \textbf{VM Uniformity} & \textbf{VM Resource Type} & \textbf{Optimization Strategy} \\ \hline
		Song et al. \cite{Song}                & Dynamic                           & Homogeneous                 & CPU                       & Heuristic                      \\ \hline
		Ni  et al. \cite{Ni}               & Static                            & Homogeneous                 & CPU \& Memory             & Heuristic                      \\ \hline
		Tordsson et al.\cite{Tordsson}          & Static                            & Heterogeneous               & Multiple                   & Meta-heuristic                 \\ \hline
		Zhao et al. \cite{Zhao}          & Dynamic                           & Homogeneous                 & CPU \& IO                 & Heuristic                      \\ \hline
		Yang et al.\cite{Yang}              & Dynamic                           & Heterogeneous               & Multiple                   & Heuristic                      \\ \hline
		Bhadani et al. \cite{Bhadani}             & Dynamic                           & Homogeneous                 & CPU                       & Heuristic                      \\ \hline
		Rouzaud-Cornabas \cite{Rouzaud}          & Dynamic                           & Heterogeneous               & CPU \& Memory             & Heuristic                      \\ \hline
		Tian et al.  \cite{Tian}            & Dynamic                           & Heterogeneous               & Multiple                   & Heuristic                      \\ \hline
		Tian and Xu \cite{Tian2}       & Static                            & Heterogeneous               & Multiple                   & Heuristic                      \\ \hline
		Thiruvenkadam et al. \cite{Thiruvenkadam}             & Dynamic                           & Heterogeneous               & Multiple                   & Hybrid                         \\ \hline
		Hu et al. \cite{Hu}                & Dynamic                           & Heterogeneous               & CPU                       & Meta-heuristic                 \\ \hline
		Wen et al. \cite{Wen}               & Static                            & Heterogeneous               & Multiple                   & Meta-heuristic                 \\ \hline
		Cho et al.  \cite{Cho}            & Dynamic                           & Heterogeneous               & Multiple                   & Meta-heuristic                 \\ \hline
	\end{tabular}
\end{table}

\begin{sidewaystable}
\caption{Algorithm Classification for Scheduling Model} 
\newcommand{\tabincell}[2]{\begin{tabular}{@{}#1@{}}#2\end{tabular}}
\centering 
\small
\begin{tabular}{|c|c|c|c|c|c|c|c|c|c|c|} 
\hline\hline 
\textbf{Algorithm} & \textbf{Scenario} & \textbf{\tabincell{c}{Experiment \\Platform}} & \textbf{Constraints}  & \textbf{\tabincell{c}{Live \\Migration} }  & \textbf{\tabincell{c}{Migration Cost \\Consideration}} &  \textbf{\tabincell{c}{Scheduling \\Objective}} & \textbf{Management}\\
\hline
　Song et al. \cite{Song} & \tabincell{c}{Public\\ Cloud} & Realistic & \tabincell{c}{Computation \&\\Communication \\Costs}   &Yes  & \tabincell{c}{Computation, \\Communication}   & \tabincell{c}{ Min \\Migration \\ Latency} & Centralized \\
\hline
　Ni et al. \cite{Ni} & \tabincell{c}{Private \\Cloud} & \tabincell{c}{Realistic \\(OpenNebula)} & \tabincell{c}{Limited \\ Resource}   &No  & \tabincell{c}{No}  & \tabincell{c}{ Min Util. SD} & Centralized \\
\hline
　Tordsson et al. \cite{Tordsson} & \tabincell{c}{Hybrid \\Cloud (Multi)} & \tabincell{c}{Realistic \\( ElasticHosts \\+ Amazon )} & \tabincell{c}{Budget,  \\User Defined}   &No   & \tabincell{c}{Computation, \\Communication }  & \tabincell{c}{ Min Costs} & Centralized \\
\hline
　Zhao et al. \cite{Zhao} & \tabincell{c}{Private\\ Cloud (Intra)} & \tabincell{c}{Realistic \\(OpenVZ)} & \tabincell{c}{Downtime}   &Yes  & \tabincell{c}{No} & \tabincell{c}{ Zero  Downtime} & Distributed \\
\hline
　Yang et al. \cite{Yang} & \tabincell{c}{Private \\Cloud} & \tabincell{c}{Simulation} & \tabincell{c}{Memory Cost\\of Migration}   &Yes   & \tabincell{c}{Memory Copy}  & \tabincell{c}{ Min Overloaded} & Centralized \\
\hline
　Bhadani et al. \cite{Bhadani} & \tabincell{c}{Public \\Cloud} & \tabincell{c}{Realistic} & \tabincell{c}{N/A}   &Yes  & \tabincell{c}{Memory, Fault\&\\ Tolerance} & \tabincell{c}{ Improve \\ Throughout} & Centralized \\
\hline
　Rouzaud-Cornabas \cite{Rouzaud} & \tabincell{c}{Public \\Cloud (P2P)} & \tabincell{c}{Simulation} & \tabincell{c}{N/A}   &Yes   & \tabincell{c}{No} & \tabincell{c}{ Faster to Solve\\ Overloaded Hosts} & Distributed \\
\hline
　Tian et al. \cite{Tian} & \tabincell{c}{Public \\Cloud} & \tabincell{c}{Simulation} & \tabincell{c}{N/A}   &Yes   & \tabincell{c}{Computation}  & \tabincell{c}{ Min Imbalance \\ Level Degree} & Centralized \\
\hline
　Tian and Xu \cite{Tian2} & \tabincell{c}{Public \\Cloud} & \tabincell{c}{Simulation} & \tabincell{c}{N/A}   & Yes   & \tabincell{c}{Computation}  & \tabincell{c}{ Min Capacity\\-makespan} & Centralized \\
\hline
　Thiruvenkadam et al. \cite{Thiruvenkadam} & \tabincell{c}{Private \\Cloud} & \tabincell{c}{Simulation} & \tabincell{c}{Overall \\ Load}   &Yes   & \tabincell{c}{Computation}  & \tabincell{c}{ Min Number \\ of Migrations} & Centralized \\

\hline
　Hu et al. \cite{Hu} & \tabincell{c}{Private \\Cloud} & \tabincell{c}{Realistic \\(OpenNebula)} & \tabincell{c}{Astringency}   &Yes   & \tabincell{c}{No} & \tabincell{c}{ Min Number \\ of Migrations} & Centralized \\
\hline
　Wen et al. \cite{Wen} & \tabincell{c}{Private \\Cloud} & \tabincell{c}{Simulation} & \tabincell{c}{Amount of \\Pheromone}   &Yes   & \tabincell{c}{Communication}  & \tabincell{c}{ Min Number of \\SLA Violations} & Distributed \\
\hline
　Cho et al. \cite{Cho} & \tabincell{c}{Private \\Cloud} & \tabincell{c}{Simulation} & \tabincell{c}{N/A}   &Yes   & \tabincell{c}{No}  & \tabincell{c}{Min Number \\ of Migrations} & Distributed \\
\hline

\hline 

\hline 
\end{tabular}
\label{tab:LPer}
\end{sidewaystable}

\subsection{Summary}
This section presents the details of the surveyed algorithms and discusses the strength and weakness of these algorithms. Table III summarizes these algorithms according to their VM　 models and Table IV assembles them based on the scheduling model. With the information, we will discuss some challenges and future work in the next section.

\section{Challenges and Future directions}
This paper investigates algorithms designed for resource scheduling in cloud computing environment. In particular, it concentrates on VM load balancing, which also refers to algorithms that balance VM placement on hosts. This paper presents classifications based on a comprehensive study on existing VM load balancing algorithms. The existing VM load balancing algorithms are analyzed and classified with the purpose of providing 
an overview of the characteristic of related algorithms. Detailed introduction and discussion of various algorithms are provided, and they aim to offer a comprehensive understanding of existing algorithms as well as further insight into the field’s future directions.

  Now we discuss the future directions and challenges as below: 

1) In the experiment platform and performance evaluation: 
\begin{itemize}
\item We see that most meta-heuristics achieve better results than traditional heuristics, while their experiments are mostly conducted under simulation toolkits. As a future direction, more meta-heuristics, like algorithms based on ACO or PSO, are encouraged to be validated under realistic platforms, which shows the possibility to implement them in real Clouds.
\item We also notice that for the VM load balancing algorithms, their optimization goals are multi-objective rather than only load balancing, such as minimizing costs or reducing downtime. Therefore, how to coordinate different optimization goals and ensure their consistency is a future research challenge.
\item Considering the diversity of our surveyed papers, we want to know which algorithm is the best or when to use which algorithm. However, these problems are still open because of the heterogeneity of different algorithms' problem formulations and lack of experiments under the same platform.
A comparative performance study for these VM load balancing algorithms under the same configuration is definitely required as future work.  

\end{itemize}

2) In the classification of VM model:
\begin{itemize}

\item  Current VM load balancing may often be dynamic, thus, a static allocation in the VM model may not be suitable. In the future,  more self-adaptive VM load balancing algorithms should be investigated. 
\item  Heterogeneous VMs are currently running in real Clouds, and CPU resource may not be the unique bottleneck, therefore, the proposed VM load balancing algorithms are preferred to be applicable for heterogeneous VMs with multiple resource in the future. 
\item  In the optimization strategy, the approach that combines heuristic and meta-heuristic is providing a promising future direction, which balances the optimized results and execution time. For example, the heuristic quickly places VMs in the initial VM placement and the meta-heuristic optimizes VM placement through VM migrations. However, how to find the balance point is a research challenge.
\end{itemize}

 3) In the classification of scheduling model: 
\begin{itemize}
\item  In cloud environment, resources are often requested concurrently and these requests may compete for resources. Our surveyed papers consider resource utilization based on current resource utilization or historic data,  while future loads are not analyzed. Thus, how to balance the VM loads considering future situation is another research challenge. 
\item  The distributed algorithms improve the system scalability and bottleneck, however, the communication cost is not discussed comprehensively and we don't know its effects on algorithm performance. Therefore, to validate the efficiency of distributed algorithms, the communication costs produced by the distributed algorithms should  also be investigated in the future. 
\item  For the algorithm designed for multiple clouds, when VMs are migrated from one cloud to another, the physical networks and virtual networks may be correlated. However, the effects under this network structure for VM migrations are not well analyzed yet, which is also another future work.
\end{itemize}

\ack This work is supported by China Scholarship Council (CSC), Australia Research Council Future Fellowship and Discovery Project Grants, National Natural Science Foundation of China (NSFC) with project ID 61672136 and 61650110513.

\bibliographystyle{wileyj}
\bibliography{ccpe}

\end{document}